\documentclass[conference]{IEEEtran}
\IEEEoverridecommandlockouts
\usepackage{url}
\usepackage{cite}
\usepackage{amsmath,amssymb,amsfonts}
\usepackage{algorithmic}
\usepackage{graphicx}
\usepackage{textcomp}
\usepackage{xcolor}
\usepackage{multirow}
\usepackage{booktabs} 
\usepackage{float}
\def\BibTeX{{\rm B\kern-.05em{\sc i\kern-.025em b}\kern-.08em
    T\kern-.1667em\lower.7ex\hbox{E}\kern-.125emX}}
\begin{document}

\title{Examining the Representation of Youth in the US Policy Documents through the Lens of Research\\
}

\author{\IEEEauthorblockN{Miftahul Jannat Mokarrama}
\IEEEauthorblockA{\textit{Computer Science} \\
\textit{Northern Illinois University}\\
Dekalb, USA \\
mmokarrama@niu.edu}
\and
\IEEEauthorblockN{ Abdul Rahman Shaikh}
\IEEEauthorblockA{\textit{Computer Science} \\
\textit{Northern Illinois University}\\
Dekalb, USA \\
ashaikh2@niu.edu}
\and
\IEEEauthorblockN{Hamed Alhoori}
\IEEEauthorblockA{\textit{Computer Science} \\
\textit{Northern Illinois University}\\
Dekalb, USA \\
alhoori@niu.edu}
}

\maketitle

\begin{abstract}
This study explores the representation of youth in US policy documents by analyzing how research on youth topics is cited within these policies. The research focuses on three key questions: identifying the frequently discussed topics in youth research that receive citations in policy documents, discerning patterns in youth research that contribute to higher citation rates in policy, and comparing the alignment between topics in youth research and those in citing policy documents. Through this analysis, the study aims to shed light on the relationship between academic research and policy formulation, highlighting areas where youth issues are effectively integrated into policy and contributing to the broader goal of enhancing youth engagement in societal decision-making processes.
\end{abstract}


\section{Introduction}
Although constituting a \textbf{substantial percentage} of the US populace, \textbf{youth} have been traditionally refused to advocate for their interests in the nation's policy-making procedures \cite{b1,b2}. Recent research conducted by Data for Progress reveals that more than two-thirds (70\%) of people aged 18 to 29 years in America perceive that their views, preferences, and ages are mostly neglected in the political realm \cite{b3}. Consequently, young people continue to have limited influence in defining their social rights with no real representational power in policymaking \cite{b1,b4}. Therefore, both policymakers and researchers agree that it is imperative to increase the engagement of children and youth in research and policy to ensure that their voice is heard \cite{b5}. However, the large volume of research conducted each year on various topics from several sources makes the analysis laborious, intricate, and challenging for researchers and policymakers \cite{b6}. In this scenario, topic Modeling, a well-known unsupervised machine learning technique, could greatly assist because of its capability to analyze extensive text data, whether structured or unstructured. With minimal processing time, it might be used to discover the underlying topics regarding youth in research and citing policies \cite{b7,b8}. Moreover, analyzing existing research and policy studies could provide stakeholders with a comprehensive idea about what topics are being discussed with respect to youth and therefore where we should focus more on increasing their representation.

In this study, our objective was to explore the research articles cited in policy documents to answer the questions below.
\begin{itemize}
    \item {\textbf{RQ1:} \textit{What topics are frequently discussed in youth research that get citations in the US policy documents?}}
    \item {\textbf{RQ2:} \textit{Is there any distinguishing pattern in the youth research topics that lead to getting more US policy citations?}}
    \item {\textbf{RQ3:} \textit{Is there any similarity or dissimilarity between topics discussed in youth research and citing US policy documents?} }
\end{itemize}
\section{Related Work}
Using topic modeling in policy analysis has also recently gained popularity among researchers. Goyal and Howlett \cite{b9} analyzed more than 13K COVID-19 policies around the world using topic modeling to examine the worldwide diversity of the COVID-19 policy guidelines and to categorize them. Craciun \cite{b10} applied \textit{LDA} topic modeling to analyze government policy documents for the internationalization of higher education. Hagen et al. \cite{b11} examined citizen-generated policy recommendations submitted through the Obama Administration's WTP petitioning system using topic modeling to facilitate the analysis of vast amounts of e-petition policies. Berliner et al. \cite{b12} classified and exposed the variety of information requests filed with Mexican federal government agencies between 2003 and 2015 using topic modeling. Bagozzi and Berliner \cite{b13} analyzed more than 6k State Department Country Reports on Human Rights Practices using topic modeling to determine the topics of interest discussed over the years on Human rights.
\section{Method}
We described our proposed topic modeling framework and workflow towards the purpose in the following subsections.
\subsection{Dataset collection}
To collect the dataset for our study, we focused on research articles that were cited within US policy documents between January 1, 2000, and December 31, 2022. The dataset was collected from \textit{Overton} \cite{b14}, the largest online repository of research articles and their corresponding citations in policy documents. The search focused on three primary topics: \textit{‘child’}, \textit{‘teen’},' \textit{‘youth’}, along with additional keywords representing various age groups related to youth, as shown in TABLE \ref{tab:keywords}, which were sourced from \textit{Related Words} \cite{b15}. For each keyword, we applied the three search criteria to maximize the coverage of relevant documents: \textit{published date}, \textit{relevance}, and \textit{citations}.

\begin{table}[htbp]
  \caption{Keywords Chosen from RelatedWords for Filtering Research Articles from Overton that Align with “Child, Teen, and Youth”}
  \label{tab:keywords}
  \centering
  \begin{tabular}{@{}lp{7cm}@{}} 
    \toprule
    \textbf{Topics} & \textbf{Keywords} \\
    \midrule
    Child & baby(-ies), kid, child(-ren), caregiver AND child, childhood, newborn, infant, toddler \\
    Teen & adolescent, adolescence, boy, girl, juvenile, teenage, teen \\
    Youth & adulthood, caregiver AND youth, youth, young \\
    Others & bully, college, foster, kindergarten, parent, preschool, school, stepchild, student \\
    \bottomrule
  \end{tabular}
\end{table}

Following the initial data collection, we filtered the dataset excluding duplicate entries, entries with missing or inconsistent publication data, and entries with unclear policy citation dates or titles. After this initial data filtration, our dataset consisted of 52,279 unique research article records cited in 35,212 policy documents with 1 to 10 policy citations. We observed that, in total, less than 1\% of research articles received more than 10 policy citations in the United States during the study period. \textbf{Since our purpose was to capture the youth topics from a vast majority of the research papers that got USA policy citations, we limited our analysis to 1 to 10 policy citations}. 

Then we checked the validity of those research articles and citing policy documents by checking the availability of their PDF documents online and propriety of the document formats. Thus we selected a total of 2301 research articles and the corresponding 2818 citing policy documents and used their titles to generate topic modeling for this research. 
This document selection process is briefly described in this research article\cite{b16}.

\begin{figure}[htbp]
    \centerline{\includegraphics[width=1\linewidth]{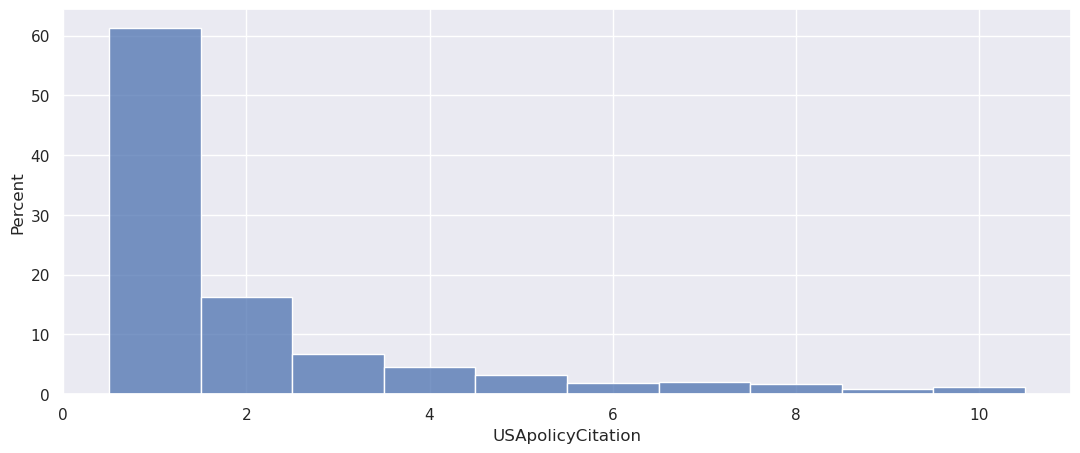}}
    \caption{Distribution of the citation count of research articles (1 to 10) in the US policy documents.}
    \label{fig:density}
\end{figure}

To generate the discussion topics of the youth research cited in the US policy, we referred to our research dataset as \textit{\textbf{research\_dataset}} and policy dataset as \textit{\textbf{policy\_dataset}}. In Fig. \ref{fig:density}, we can notice the citation count disparity between counts 1 and 2 to 10. This indicates that a large portion of the research articles got only one policy citation, however, getting more than one citation is not a frequent phenomenon in the policy domain. In this research, we are interested to see the reason for such disparity through a topic modeling and analysis approach. Therefore, we further divided our \textit{\textbf{research dataset}} into the following two subsets.
\begin{itemize}
    \item {\textit{\textbf{research\_dataset\_1}}: Research articles that got a citation in only one policy document.}
    \item {\textit{\textbf{research\_dataset\_0}}: Research articles that got citations in 2 to 10 policy documents.}
\end{itemize}

\subsection{Topic Modeling}

\begin{figure}[htbp]
    \centering
    \includegraphics[width=0.90\linewidth]{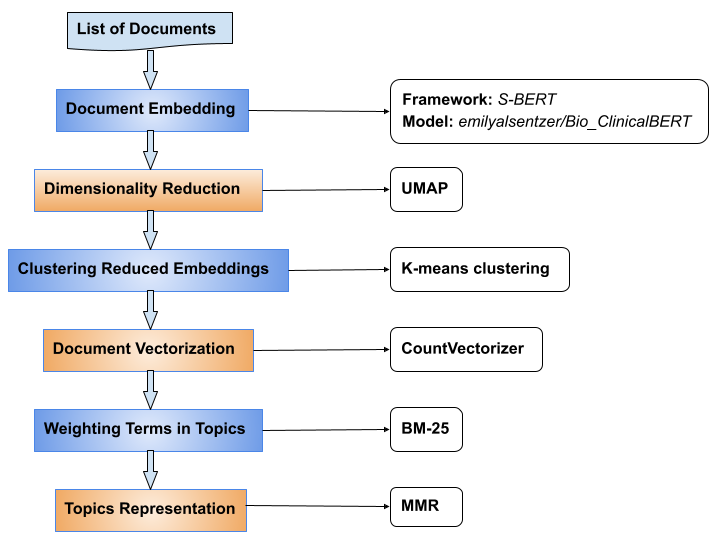}
    \caption{BERTopic framework used for topics generation.}
    \label{fig:bertopic}
\end{figure} 
We used \textit{BERTopic}, a transformer-based topic modeling framework \cite{b17}, to generate topics for each subset. There were a total of six modules that were used sequentially to build our BERTopic framework, as shown in Fig. \ref{fig:bertopic}. We set the value of \textit{top\_n\_words} to 15 while building the topic model in this framework. This parameter represents the number of words that would be returned by each topic in the model. The modules are tuned with the corresponding hyperparameters as described in the following.
\begin{enumerate}
    \item \textbf{Document Embedding:} In this module, we generated embeddings of the text list where each text represents each pre-processed research or policy text file. We used the sentence-BERT framework with the pretrained large language model \textit{`Bio\_ClinicalBERT'} \cite{b18} for this purpose, generating 768-dimensional embeddings for each input text corresponding to the policy or research article. 
    \item \textbf{Dimensionality Reduction:} In this module, the dimension of each text is reduced using \textit{Uniform Manifold Approximation and Projection} (\textit{UMAP}). UMAP is a non-linear method that provides a more meaningful representation of complex, non-linear data at the cost of computational efficiency and less direct interpretability. We used the following set of parameter values for \textit{UMAP} for each data subset: \\
    \textit{n\_neighbors} = 30,\textit{ n\_components} = 3, \textit{min\_dist} = 0.00, and \textit{metric} = 'cosine'\\
    
    \item \textbf{Clustering Reduced Embeddings:} In this module, we clustered the reduced embeddings into topics using the \textit{K-means} clustering technique. \textit{K-means} clustering is an unsupervised machine learning algorithm used for partitioning a set of data points into a specified number of clusters \cite{b19}. It is a pretty straightforward algorithm and works best for our dataset in the given experimental settings. In our experiments, we set the value of \textit{n\_clusters} to 15. 
    
    \item \textbf{Document Vectorization:} We used scikit-learn \textit{CountVectorizer}
    to tokenize each text in the cluster. We set the \textit{ngram\_range} values between 1 and 3 and removed updated stop words from the text. Since it is a count-based method, we updated the by default stop word list, including alphabets, words with length two in \textit{NLTK} vocabulary, and a list of custom words ([`',` ',`et',`cox',`md',`phd',`ms',`mphil',`mph']) that frequently appear, producing no meaningful topics.
    \item \textbf{Weighting Terms in Topics:} We used the class-based BM-25
    weighting technique to calculate and rank the relevance of the terms in topics as it provides better results for the extensive and diverse collection of datasets with variable-length texts. Additionally, to penalize the most frequent words not listed in the stop word list during ranking, we set the parameter \textit{reduce\_frequent\_words} to `True.'
     \item \textbf{Topics Representation:} Finally, to reduce the repetitiveness of similar terms (e.g., diseases, disease) and enhance the variety of keywords, we further fine-tuned the representation of the topics using the Maximal Marginal Relevance (MMR) algorithm. We set a \textit{diversity} score of 0.8, leading to choosing keywords that maximize their diversity within the document.
 \end{enumerate}
\begin{figure}[htbp]            
   \centerline{\includegraphics[width=0.90\linewidth]{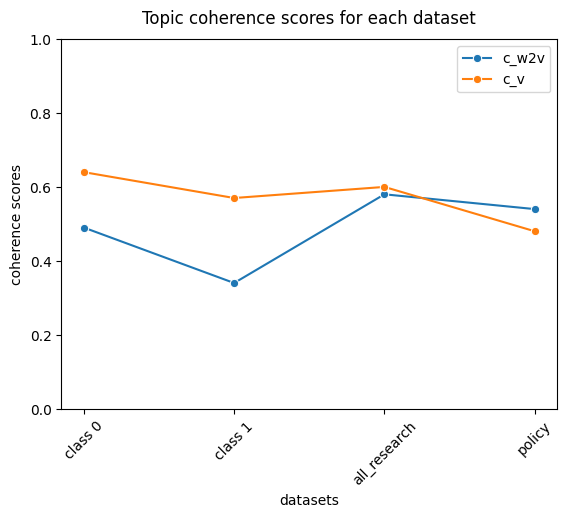}}
    \caption{Topic coherence of the topics generated in each dataset.}
    \label{fig:coherence}
\end{figure}
In Fig. \ref{fig:coherence}, the coherence scores of each topic model generated from four categories of datasets (\textit{\textbf{research\_dataset}}, \textit{\textbf{policy\_dataset}}, \textit{\textbf{research\_dataset\_1}}, \textit{\textbf{research\_dataset\_0}}) are shown. 

We used two topic coherence metrics: (1) \textit{C\_V} coherence metric, which measures pointwise mutual information (PMI) between words in a topic, and (2) \textit{C\_W2V} coherence metric, which uses Word2Vec word embeddings to measure the semantic similarity of words in a topic. However, it is worth noting that the interpretability and relatedness of the topics are more dependent on domain knowledge \cite{b20}, and therefore, we also did a visual inspection of the topics generated to evaluate their topic coherence. 

\section{Topic Analysis}
From the Fig. \ref{fig:coherence}, we noticed that youth topics generated from \textbf{\textit{research\_dataset\_1}} (class 1) are more versatile than those in \textbf{\textit{research\_dataset\_0}} (class 0). On the other hand, when the research dataset is not classified (all\_research), the generated topics exhibit more coherence and relatedness and less diversity. Overall, both the research dataset and the policy dataset exhibited a moderate level of diversity, with the research dataset demonstrating slightly higher coherence scores. 

To answer the \textbf{RQ1}, from the topics generated, we broadly noticed five broad categories of topics being discussed in the research cited in the US policy documents on youths. They are listed in TABLE \ref{tab:topics}. We observed that youth research topics that get coverage in policy
are mostly related to healthcare.

\begin{table}[htbp]
\caption{All topics generated from \textit{research\_dataset}}
\begin{center}
\begin{tabular}{@{}lp{5.5cm}p{2.10cm}@{}}
\hline
{\textbf{Sl No.}} & \textbf{Broader Topics} & \textbf{Percentage in Research Articles} \\
    \hline
    1. & Clinical experiments and research in healthcare and medicine & Around 26\%\\
    2. & Physical and psychological issues affecting proper youth development & Around 22\%\\
    3. & Global climate and health issues that affect the youth lifestyle and well-being & Around 22\%\\
    4. & Early vaccination targeting preventable diseases and maternal and infant health issues & Around 20\%\\
    5. & Highly transmitting global pandemic (e.g., COVID-19, SARS) for which policymakers need urgent research evidence & Around 11\%\\
  \hline
\end{tabular}
\label{tab:topics}
\end{center}
\end{table}

We used the topics generated from \textit{\textbf{research\_dataset\_0}} and \textit{\textbf{research\_dataset\_1}} to answer \textbf{RQ2}. We found that research that got more citations (\textit{\textbf{research\_dataset\_0}}) focused more on topics related to the global COVID-19 pandemic and its impact on youth well-being (around 51\%). However, the scenario differs from the one in research that got only one citation (\textit{\textbf{research\_dataset\_1}}). For example, only about 7\% of the research articles are identified as discussing the topic of the COVID-19 pandemic on a broader scale in this case. Moreover, research with more policy citations aligns more with \textit{topic category 2} and \textit{topic category 4}. However, in the case of research articles with \textbf{one} citation only, the topics are more versatile and reflect all the topic categories listed in TABLE \ref{tab:topics}. 

\begin{figure}
    \centering
    \includegraphics[width=1\linewidth]{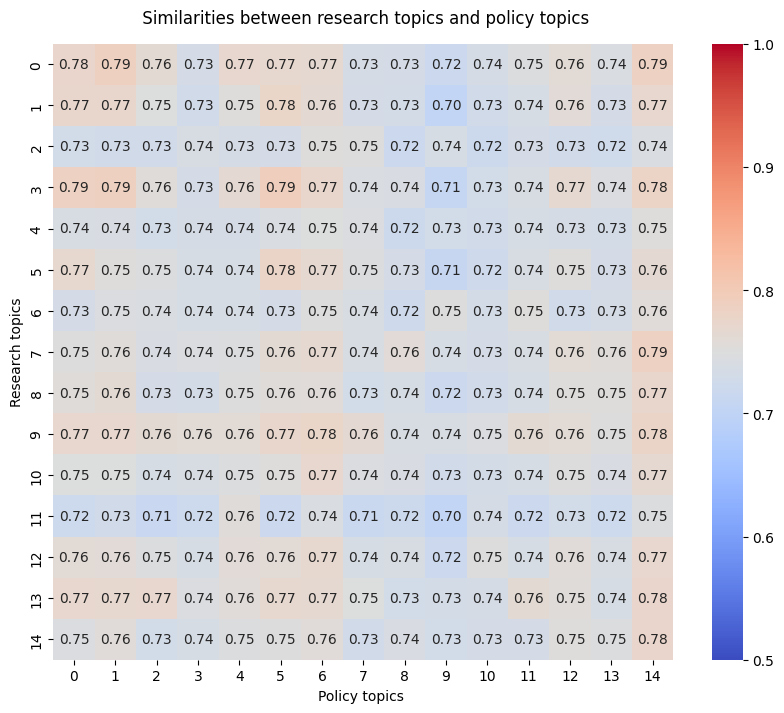}
    \caption{Similarities between research and policy topics}
    \label{fig:researc_policy_sim}
\end{figure}

We used the topics generated from \textit{\textbf{research\_dataset}} and the topics generated from \textit{\textbf{policy\_dataset}} to answer our last research question, \textbf{RQ3}. We found some topics frequently appearing in policy documents, representing the official terms used while policy documenting (e.g., report, microsoft word) and places (e.g., New York, Washington) where policymaking institutions are located. Besides these, the topics discussed in the policy documents are almost similar to those discussed in the research articles, as we categorized in TABLE \ref{tab:topics}. We calculated the \textit{cosine} similarity scores between topics in the \textit{\textbf{research\_dataset}} and \textit{\textbf{policy\_dataset}} using the same Large Language Model used for topic modeling in this experiment [\textit{Bio\_ClinicalBert} (Fig. \ref{fig:researc_policy_sim})]. We noticed more than 70\% similarities between the topics that align with our visual inspection and coherence scores.

The data and code used for this work can be accessed through this link: \url{https://github.com/JannatMokarrama07/Research-Policy-Topic-Modeling}.
\section{Disucssion}
Organizations can improve the performance and accuracy of their AI models by leveraging domain-specific datasets \cite{b21}. Bio\_ClinicalBERT, a specialized variant of Google’s BERT, was trained with parameters initialized from BioBERT (BioBERT-Base v1.0 + PubMed 200K + PMC 270K) using code from the original BERT repository. This model was further pretrained on the MIMIC-III database, which comprises electronic health records from ICU patients at Beth Israel Deaconess Medical Center in Boston, USA \cite{b22}. In our previous study \cite{b16}, we found that Bio\_ClinicalBERT provides a better embedding for our dataset and research context compared to other pretrained models like Scibert \cite{b23}. Therefore, we used this pre-trained large language model for better domain-specific topic modeling. However, the limited data collection and reliance on the Overton repository impose constraints on our experiments. In future, we plan to exapnd our dataset beyond Overton using \textit{Web of Science} \cite{b24}, \textit{Altmetric} \cite{b25}, and \textit{Dimension} \cite{b26} online databases to cover data from diverse research and policy sources. 
\section{Conclusion}
In this research, we investigated the prevalence of topics in youth research articles and US policy documents using the state-of-the-art transformer-based topic modeling technique BERTopic. The experiment was done to identify whether research articles with specific topics are getting more attention than others. Moreover, finding out the existence of topic similarities (or dissimilarities) between cited research articles and citing policy documents was also a crucial part of this experiment. This research will be helpful for researchers to understand what research topics get attention among policymakers and, therefore, investigate whether policymakers need to focus on other topics that are, in general, overlooked. It can also assist future researchers in further inspecting the factors that hinder their research topics from garnering proper attention to policy compared to other topics. In the future, we plan to increase the dataset size and extend the work by experimenting with different LLMs and comparing them with the traditional \textit{LDA} topic modeling. We believe that our experiment will be useful for both policymakers and researchers to focus on new research directions and make their decisions for the youth benefits.

\section*{Acknowledgment}
We are grateful to Terry Bucknell and his team to give us access to \textit{Overton} dataset that was invaluable support for this research.
 






\end{document}